\documentclass{INTERSPEECH2023}

\interspeechcameraready 
\usepackage{multirow}
\usepackage{multicol}
\usepackage{caption}
\usepackage{cite}

\title{DeepVQE: Real Time Deep Voice Quality Enhancement for Joint Acoustic Echo Cancellation, Noise Suppression and Dereverberation}
\name{Evgenii Indenbom, Nicolae-C\u{a}t\u{a}lin Ristea, Ando Saabas, Tanel P\"arnamaa, Jegor Gužvin, Ross Cutler}

\address{
  Microsoft Corp.
}
\email{ando.saabas@microsoft.com}

\begin{document}

\maketitle

\begin{abstract}
Acoustic echo cancellation (AEC), noise suppression (NS) and dereverberation (DR) are an integral part of modern full-duplex communication systems. As the demand for teleconferencing systems increases, addressing these tasks is required for an effective and efficient online meeting experience. Most prior research proposes solutions for these tasks separately, combining them with digital signal processing (DSP) based components, resulting in complex pipelines that are often impractical to deploy in real-world applications. This paper proposes a real-time cross-attention deep model, named DeepVQE, based on residual convolutional neural networks (CNNs) and recurrent neural networks (RNNs) to simultaneously address AEC, NS, and DR. We conduct several ablation studies to analyze the contributions of different components of our model to the overall performance. DeepVQE achieves state-of-the-art performance on non-personalized tracks from the ICASSP 2023 Acoustic Echo Cancellation Challenge and ICASSP 2023 Deep Noise Suppression Challenge test sets\footnote{This paper was not entered in these ICASSP 2023 challenges}, showing that a single model can handle multiple tasks with excellent performance. Moreover, the model runs in real-time and has been successfully tested for the Microsoft Teams platform.

\end{abstract}

\vspace{-0.25cm}
\section{Introduction}
Teleconferencing systems like Microsoft Teams, Skype, and Zoom have seen a surge in demand due to the rise of remote work in fields such as business, education, and healthcare. To ensure that such systems provide a productive and pleasant experience to users, it is crucial that they provide good call quality. Noise and acoustic echo are among the main causes of call quality degradation that can significantly diminish speech intelligibility and hinder communication \cite{grancharov2008speech}. Those problems become even more challenging in full duplex communication when echo interferes with double-talk (DT) scenarios \cite{sridhar2021icassp}. Therefore, solutions that can address acoustic echo, noise, and dereverberation are essential for enabling seamless communication.

Although acoustic echo, noise, and reverberation are theoretically three separate effects, they are interdependent in real communication systems, which can make it challenging to address them individually. For instance, in high noise or reverberant scenarios, the performance of the AEC system is negatively affected. Most related works have focused on each task separately \cite{zhang2022multi, zheng2021interactive, zhang2022deep, indenbom2022deep, ju2022tea, zhang2022multitask, soni2023state, zhang2021deep, westhausen2020dual, zheng2022time}. However, such an approach involves cascading the AEC, NS and DR systems, which leads to more complex communication pipelines. To address this shortcoming, we propose a more natural approach by canceling echoes, noise, and reverb with a joint deep learning model.


Recently, joint AEC and NS \cite{xu2022deep, chen2022fullsubnet, zhang2022multi, wang2022nn3a} methods have been developed to simplify the communication pipeline while providing a good AEC and NS performance. For example, MTFAA-Net \cite{zhang2022multi} is a neural network for joint AEC and NS, based on multi-scale time-frequency processing and streaming axial attention. The network provided the best results in 2022 AEC and DNS challenges, showing that the model can handle joint tasks with state-of-the-art (SOTA) performance. Still, MTFAA-Net  relies on classical AEC components, such as a signal processing-based linear echo canceller (LAEC) and a delay compensator for aligning the microphone and the far end signals. In fact, in the ICASSP 2022 AEC Challenge \cite{cutler2022icassp}, the top 4 submissions employed a DSP-based LAEC and a delay compensator for signal alignment \cite{zhang2022multi,deep2022zhao,multi2022zhang,explore2022xingwei}. We show a non-hybrid approach can give SOTA performance. 

The methods \cite{indenbom2022deep, ma2021echofilter, liu2022sca} perform in-model alignment between the microphone and far end signals, which can replace and improve DSP-based alignment methods \cite{ianniello1982time}. In \cite{ma2021echofilter} a model which performs alignment in the time domain with a local attention block is described, where the attention block computes the alignment based on the RNN's internal states. Distinctly, our model performs attention to deep time-frequency features, computing an actual delay distribution which is used to soft align the far end features. More closely to our work, \cite{indenbom2022deep} uses a cross-attention mechanism for deep features alignment. We enhance that mechanism by adding a convolutional layer in the time-delay map, which stabilizes the delay distribution and enhances the AEC performance. There has been a wide range of NS methods (e.g., \cite{zheng2021interactive, xu2022deep, chen2022fullsubnet}) with remarkable results, but without the ability to jointly perform AEC.

\begin{figure*}[t]
  \centering
  \includegraphics[width=\linewidth]{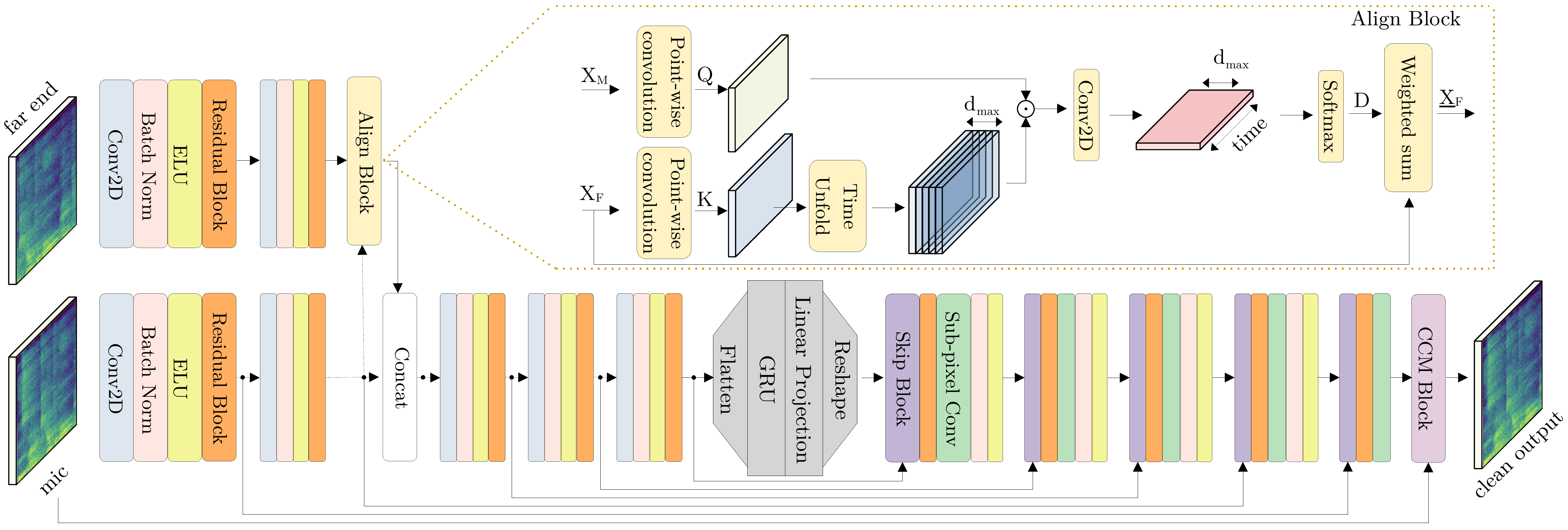}
  \caption{DeepVQE architecture overview. Using the mic and far end signal, the network reconstructs the clean mic signal, without undesired echoes, noises, and reverb. At the top, the cross-attention-based alignment block is illustrated. Best viewed in color.}
  \label{fig_deepvqe}
  \vspace{-0.4cm}
\end{figure*}

In this paper we describe DeepVQE, a residual CNN auto-encoder with a Gated Recurrent Unit \cite{chung2014empirical} (GRU) bottleneck. Considering that one major issue in AEC systems is the delay between the microphone and the far end reference signals \cite{indenbom2022deep}, our network contains a cross-attention block to soft-align the near end and far end signals. To provide a better output signal quality, we employed a complex convolving mask (CCM) block, being inspired by \cite{mack2019deep}, which allows the network to estimate each time-frequency bin by mixing multiple neighbor bins in a learnable fashion. Moreover, being inspired by the success in the computer vision field, we replaced the standard upscaling blocks from the decoder with sub-pixel convolutional blocks \cite{shi2016real}, enabling a higher feature diversity with a small performance cost.  Our contribution are:
\begin{itemize}
    \item We developed a new cross-attention mechanism for the microphone and far end soft alignment in feature space.
    \item We developed a new architecture which efficiently combines multiple ideas (alignment block, residual block, CCM, sub-pixel convolution) for joint AEC, NS and DR tasks.
    \item We obtained state-of-the-art performance on both Acoustic Echo Cancellation Challenge 2023\cite{AEC-2023} and Deep Noise Suppression Challenge 2023 \cite{DNS-2023} with our joint model. 
    \item DeepVQE-S joint model is performant enough for real-time workloads on even low end devices and has been successfully tested in Microsoft Teams for hundreds of millions of users.
\end{itemize}

\vspace{-0.2cm}
\section{Proposed Method}

\noindent \textbf{Problem Formulation.}
We consider the following generic communication system for AEC, NS, and DR tasks: the far end reference signal is transmitted to the receiving room, played back through the loudspeaker, and then picked up by the microphone via an acoustic echo path (modeled by a room impulse response). The captured microphone signal is composed of near end signal, background noise, reverberations, and echoes. The captured microphone signal is then processed by the speech enhancement component, and the produced clean signal is sent to the far end user.
The speech enhancement component's responsibility is to perform echo cancellation, as well as to remove background noises and reverberation from the near end signal. Our goal is to provide a single model performing the joint task of removing undesired echoes, noises, and reverberation, having the microphone and the reference far end signals as inputs.

\noindent \textbf{Feature extraction.}
\label{feat_extraction}
Because there is not a significant perceptual difference between fullband (48 kHz) and super wideband (24 kHz) signals \cite{beerends_subjective_2020}, we chose to sample input and output audio at 24 kHz to achieve better inference speed. The 48 kHz signals are downsampled, enhanced with our models, and upsampled back to the original sampling rate. The preprocessing is performed identically for the reference far end and microphone signals. The input features to the network are power law compressed complex spectra computed with a squared root Hann window \cite{braun2021towards}.

\noindent \textbf{Overall network architecture.}
The DeepVQE architecture is depicted in Figure \ref{fig_deepvqe}. It consists of encoder, GRU bottleneck, decoder and CCM block, which we describe below.
In this section we use $c, t, f \in \mathbb{N}$ to denote channel, time, and frequency axis lengths.

\noindent \textbf{Encoder.}
The encoder is composed of the mic and far end branches. The microphone branch has five encoding blocks, while the far end branch has only two, followed by the alignment block. The alignment block aligns far end and microphone features in time. The aligned far end and microphone features are concatenated and fed into the third encoding block in the microphone branch. Each encoding block is built by stacking a downsampling convolutional layer, batch-norm, ELU function, and residual block (see below). The first microphone encoding block has 64 filters and all the following microphone encoding blocks have 128 filters. The far end branch has 32 filters in the first block and 128 filters in the second. The downsampling convolutional layers have kernel size $4 \times 3$ and a stride of $1 \times 2$, reducing the number of bins along the frequency axis. The convolutions are causal, meaning that the padding is performed so that no look-ahead is used. The encoder is shown in the left part of Figure \ref{fig_deepvqe}.

\noindent \textbf{Residual block.}
In both the encoding and decoding blocks we added a residual block, which increases the network's capacity, while not hindering the gradient flow through the network. The block consists of a convolutional layer, followed by batch-norm and exponential linear unit (ELU) \cite{clevert2015fast} activation. The block could be formally defined as:
\vspace{-0.4cm}

\begin{equation}
\vspace{-0.1cm}
    \textbf{Y} = \textbf{X} + ELU( BatchNorm (Conv2D(\textbf{X}))),
\end{equation}

\noindent
where $\textbf{X}$ and $\textbf{Y} \in \mathbb{R}^{c \times t \times f}$ are input and output tensors respectively. The convolution layer in the residual block has the same number of filters as the number of channels in the input, kernel size of $4 \times 3$, the stride of $1$, and causal padding such as the input shape is preserved.

\noindent \textbf{Alignment block.}
Let $\textbf{X}_M\in \mathbb{R}^{c \times t \times f}$ be the mic features and $\textbf{X}_F\in \mathbb{R}^{c \times t \times f}$ the far end features. 
The feature maps are processed by point-wise convolution layers into $\textbf{Q}\in \mathbb{R}^{h \times t \times f}$ and $\textbf{K}\in \mathbb{R}^{h \times t \times f}$, where $h$ is the number of similarity channels. Next, we unfold on the time axis the $\textbf{K}$, creating a delay dimension and changing the shape to $\textbf{K}_{u}\in \mathbb{R}^{h \times t \times d_{max} \times f}$, where $d_{max}$ is the maximum echo delay expressed in time frames. Afterward, we perform a dot product on the frequency axis between the query and the unfolded key, obtaining $\textbf{Z}\in \mathbb{R}^{h \times t \times d_{max}}$. The results are fed into a convolutional layer with a kernel size of $5 \times 3$, padding of $3 \times 1$, and stride 1. The convolution has a single filter combining $h$ similarity channels into a single attention head, which is further processed by a softmax on the delay axis, outputting a delay probability distribution $\textbf{D}\in \mathbb{R}^{t \times d_{max}}$. Finally, the aligned far end features $\underline{\textbf{X}}_F\in \mathbb{R}^{c \times t \times f}$ are computed as a weighted sum on the time axis with the corresponding delay probabilities from $\textbf{D}$. More precisely, for each delay value in $[0, d_{max})$, $\textbf{X}_F$ is delayed, multiplied by the corresponding weight factor from $\textbf{D}$, and added to the final result $\underline{\textbf{X}}_F$.

\noindent \textbf{Bottleneck.}
The bottleneck is located between the encoder and decoder and consists of a recurrent layer and a linear projection. The recurrent layer input is feature maps from the encoder flattened along the channel and frequency dimensions. Formally, the input $\textbf{X}\in \mathbb{R}^{c \times t \times f}$ is flattened into $\textbf{X} \in \mathbb{R}^{t \times (c \cdot f)}$. Afterward, $\textbf{X}$ is processed into the recurrent layer, fed into the linear projection, and then the linear projection output is reshaped back to $\textbf{X}\in \mathbb{R}^{c \times t \times f}$. Following \cite{braun2021towards}, considering that an LSTM does not bring significant performance improvements, we use a GRU layer to reduce the model complexity. Using linear projection after the recurrent layer allows for a reduction of the number of hidden units in the recurrent layer improving both performance and training stability.

\noindent \textbf{Decoder.}
The decoder consists of five decoding blocks. All but the last one is built by stacking a skip block, residual block, sub-pixel convolution block, batch-norm, and ELU function. The last decoding block consists of a skip block, residual block, and sub-pixel convolution block only. The number of filters in the decoder is changed only in the sub-pixel convolution blocks, while the other blocks preserve the tensor shape. The number of filters for the sub-pixel blocks is 128, 128, 128, 64, and 27. Each sub-pixel convolution has the kernel size of $4 \times 3$ and a stride of 1. All convolutions are causal, meaning that the padding is added so that no look-ahead is performed.

\noindent \textbf{Skip block.}
We replaced the classical skip connection, based on concatenation or summing, with a convolutional layer, having a kernel size of $1 \times 1$ and a stride of $1$. The encoder features are point-wise projected and then summed with the corresponding decoder output. Besides decoupling the encoder and decoder feature spaces, the point-wise convolution allows us to choose the number of channels in the encoder independently from the number of channels in the decoder and obtain better results for the performance-speed trade-off.

\noindent \textbf{Sub-pixel convolution.}
After downscaling the input on the frequency axis in the encoder part, we need to upscale back to the original resolution in the decoder. We replaced the regular upscaling method based on transposed convolution with the sub-pixel convolution \cite{shi2016real}, which learns an array of filters to upscale the low-resolution feature maps into the high-resolution output. Each upscaling is performed with a factor of two on the frequency axis.
Formally, the $\textbf{X}\in \mathbb{R}^{c_i \times t \times f}$, having $c_i \in \mathbb{N}$  channels, is transformed by a regular convolution with $2c$ filters into $\textbf{X}'\in \mathbb{R}^{2c \times t \times f}$, and then transposed and reshaped into the actual output $\textbf{Y}\in \mathbb{R}^{c \times t \times 2f}$.

\noindent \textbf{Complex convolving mask block.}
The complex convolving mask block consists of two stages. The first stage builds the complex-valued mask by splitting the input channels into three weight components. Each component is the weight of a $120$ degree rotating vector in the complex plane. We define $\textbf{v} = (v_1, v_2, v_3) = (1, -\frac{1}{2} + j\frac{\sqrt{3}}{2}, -\frac{1}{2} - j\frac{\sqrt{3}}{2})$ and reshape the input $\textbf{X}\in \mathbb{R}^{c \times t \times f}$ into  $\textbf{X}'\in \mathbb{R}^{3 \times \frac{c}{3} \times t \times f}$. Next, we compute the complex mask $\textbf{H} \in \mathbb{C}^{\frac{c}{3} \times t \times f}$ as described in Equation \ref{eq_ccm}.
\vspace{-0.3cm}
\begin{equation}
\vspace{-0.15cm}
    \textbf{H} = \textbf{v} \cdot \textbf{X}'
\label{eq_ccm}
\end{equation}

\noindent
Considering that the angle between $\textbf{v}$ components is $120$ degrees in the complex space, the complex mask covers the entire complex plane. In practice we observed that using a three-vector component instead of the regular two-vector component (real and imaginary parts) offers more stable output results, preventing low noise and echo leakage.

In the second stage, we reshape the channel dimension of the complex mask $\textbf{H}$ to form a $(m +1) \times (2n+1)$ convolution kernel $\textbf{M} \in \mathbb{C}^{(m+1) \times (2n+1) \times \frac{c}{3(m+1)(2n+1)} \times t \times f}$ with weights varying over time and frequency dimensions. But, as the input microphone spectrum is a single channel complex valued tensor $\textbf{X}_{mic} \in \mathbb{C}^{ t \times f}$, we need to enforce $c = 3(m+1)(2n+1)$. Therefore, having the input microphone spectrum $\textbf{X}_{mic} \in \mathbb{C}^{ t \times f}$ and the complex convolving mask $\textbf{M} \in \mathbb{C}^{(m + 1) \times (2n + 1) \times t \times f}$ (after squeezing the redundant channel dimension), the clean spectrum $\hat{\textbf{X}} \in \mathbb{C}^{ t \times f}$ is estimated as described in Equation \ref{eq:X}. The input spectrum $\textbf{X}_{mic}$ is padded with zeros to ensure that clean spectrum $\hat{\textbf{X}}$ is produced for all frames and frequency bins.
\vspace{-0.2cm}
\begin{equation}
\label{eq:X}
\vspace{-0.2cm}
    \hat{\textbf{X}}(t, f) = \sum_{i=-m}^{0} \sum_{j=-n}^{n} \textbf{X}(t + i, f + j) \cdot \textbf{M}(i, j, t, f)
\end{equation}

Computing a deep filter for output reconstruction helps the network to leverage neighbor time-frequency bins in a learnable fashion. The CCM block is applied causally.

\vspace{-0.2cm}
\begin{table*}[!t]
\caption{Ablation study results for our small DeepVQE-S model on LD-M, LD-H \cite{indenbom2022deep}, and 2023 AEC challenge \cite{AEC-2023} blind test set, for far end single-talk (AEC-FEST) and double-talk (AEC-DT) scenarios. We compare the DeepVQE-S model without an alignment block (shown as DSP-aligned), with the alignment block proposed in \cite{indenbom2022deep} and with our alignment block (shown as Ours). For the model without the alignment block, the far end and mic signals are aligned using the DSP-based method. We report the ERLE, WER, AECMOS Echo (AECMOS$_e$) and Degradation (AECMOS$_d$) \cite{purin2022aecmos}.}
\label{tab_results_align}
\setlength\tabcolsep{3.5pt}
\centering
\begin{tabular}{ c | c c | c c | c c | c c c }

{Align} &
\multicolumn{2}{c|}{{LD-M}} &
\multicolumn{2}{c|}{{LD-H}} &
\multicolumn{2}{c|}{{AEC-FEST}} &
\multicolumn{3}{c}{{AEC-DT}} \\
 {method} & {ERLE$\uparrow$} & 
  {AECMOS$_e$$\uparrow$} &  {ERLE$\uparrow$} &{AECMOS$_e$$\uparrow$}  &  {ERLE$\uparrow$} &{AECMOS$_e$$\uparrow$} &   {AECMOS$_e$$\uparrow$} & {AECMOS$_d$$\uparrow$} & {WER$\downarrow$} \\
\hline

DSP-aligned & 41.76 & 4.15 & 33.18 & 3.96 &  54.12 & 4.45 & \textbf{4.62} & 3.89 & 36.27\\

\cite{indenbom2022deep} & 59.19 & 4.57 & 54.49 & 4.46  & 61.04 & 4.56 & 4.60 & 3.95 & 36.23\\

Ours & \textbf{61.22} & \textbf{4.60} & \textbf{55.51} & \textbf{4.49}  & \textbf{65.70} & \textbf{4.61} & \textbf{4.62} & \textbf{4.02} & \textbf{31.79}\\

\hline
\end{tabular}
\vspace{-0.1cm}
\end{table*}

\begin{table*}[!t]
\caption{Ablation study results for our small DeepVQE-S model on 2023 AEC Challenge \cite{AEC-2023} near end single-talk (AEC-NEST) blind test set and 2022 DNS Challenge \cite{dubey2022icassp} blind test set. The model includes our alignment block although the far end signal is empty in nearend scenario. We report the AECMOS Degradation (AECMOS$_d$) \cite{purin2022aecmos}, DNSMOS P.835 SIG, BAK, OVRL scores \cite{reddy2022dnsmos} and the SRR, estimated with an internal model. On the first line, we included for comparison the scores for noisy data.}
\label{tab_results_ns}
\setlength\tabcolsep{3.5pt}
\centering
\begin{tabular}{ c| c| c | c c c c c | c c c c c}

{Residual} & {Sub-Pixel} & {CCM} &
\multicolumn{5}{c|}{{AEC-NEST}} &
\multicolumn{5}{c}{{DNS 2022}} \\
 {Block} & {conv} & {Block} &
{AECMOS$_d$$\uparrow$} & {SIG$\uparrow$}  &  {BAK$\uparrow$} & {OVRL$\uparrow$} & {SRR$\uparrow$} & {AECMOS$_d$$\uparrow$} & {SIG$\uparrow$} & {BAK$\uparrow$} & {OVRL$\uparrow$} & {SRR$\uparrow$}\\
\hline


\hline
 -& - &- & 3.27 & 3.70 & 3.08 & 2.90 & 25.18 & 2.72 & 3.52 & 2.10 & 2.28 & 25.56 \\
\hline
             &  $\checkmark$ &  $\checkmark$& 4.35 & 3.82 & 4.30 & 3.52 & 35.83 &4.04 & 3.57 & 4.06 & 3.26 & 36.71\\
  $\checkmark$ &               &  $\checkmark$& 4.35 & 3.83 & 4.32 & 3.55 & 35.62 & 4.07 & 3.58 & 4.03 & 3.26 & 36.45\\
 $\checkmark$ &  $\checkmark$ &              &  4.29 & 3.80 & 4.30 & 3.51 & 34.65 & 3.95 & 3.54 & 4.02 & 3.22 & 35.71\\
 $\checkmark$ &  $\checkmark$ &  $\checkmark$& \textbf{4.36} & \textbf{3.84} & \textbf{4.35} & \textbf{3.56} & \textbf{36.27} &\textbf{4.09} & \textbf{3.60} & \textbf{4.10} & \textbf{3.30} & \textbf{36.98}\\
\hline
\end{tabular}
\vspace{-0.4cm}
\end{table*}

\begin{table}[!t]
\caption{2023 DNS Challenge MOS results \cite{DNS-2023}. We included the competition baseline and the non-personalized winner.}
\label{tab_results_DNSmos}
\centering
\setlength\tabcolsep{4.0pt}
\begin{tabular}{c | l | c c c c |c}
 & {Method} & {SIG} & {BAK} & {OVRL} & {WAcc} & M\\ 
\hline
{\multirow{3}{*}{\rotatebox[origin=c]{90}{Track 1}}} 
& DNS Baseline   & 3.14 & 2.60 & 2.34 & 70.7\% & 0.521 \\
& DNS Winner \cite{DNS-2023} & \textbf{3.58} & 2.82 & 2.65 & 72.5\% & 0.569 \\
& DeepVQE  & 3.47 & \textbf{2.94} & \textbf{2.73} & \textbf{73.4\%} & \textbf{0.582}\\
\hline
{\multirow{3}{*}{\rotatebox[origin=c]{90}{Track 2}}} 
& DNS Baseline    & 3.22 & 2.68 & 2.38 & 72.7\% & 0.536 \\
& DNS Winner \cite{DNS-2023} & \textbf{3.64} & 2.88 & 2.66 & 72.4\% & 0.570 \\
& DeepVQE  & 3.57 & \textbf{3.06} & \textbf{2.83} & \textbf{76.0\%} & \textbf{0.608}\\
\hline
\end{tabular}
\vspace{-0.4cm}
\end{table}

\begin{table}[!t]
\caption{2023 AEC Challenge MOS results \cite{AEC-2023}. We included the competition baseline and the non-personalized winner.}
\label{tab_results_AECmos}
\centering
\setlength\tabcolsep{1.5pt}
\begin{tabular}{l | c c c c c c | c}
 \multirow{2}{*}{Method} & {ST FE} &  {DT} & {DT} & {ST NE} &  {ST NE} & \multirow{2}{*}{WAcc} &  Final \\ 
 & Echo & Echo &  Other  & SIG & BAK & & score \\

\hline
AEC Baseline    & 4.53 & 4.28 & 3.47 & 3.88 & 3.88 & 64.9\%  & 0.736\\
AEC winner    &  \textbf{4.70} & \textbf{4.77} & \textbf{4.31} & 3.99 & 4.38 & \textbf{82.3 \%} & 0.852 \\
DeepVQE-S    &  4.66 & 4.63 & 4.00 & 4.04 & 4.33 & 75.7\% & 0.821\\
DeepVQE  &  4.69 & 4.70 & 4.29 & \textbf{4.15} & \textbf{4.41} & 80.7\% & \textbf{0.854}\\

\hline
\end{tabular}
\vspace{-0.5cm}
\end{table}

\section{Experiments}

\noindent \textbf{Datasets.}
To ensure the generalization capacity, the training data are synthesized online from clean and noisy speech, with random parameters for each sample (e.g., signal-to-noise ratio, room impulse response, distortion, gain, signal-to-echo ratio etc.). We sample clean and noisy speech, and noise recordings for training from data provided in the ICASSP 2022 AEC \cite{cutler2022icassp} and DNS \cite{dubey2022icassp} challenges.

We report the final results on the blind test sets from the ICASSP 2023 AEC \cite{AEC-2023} and DNS \cite{DNS-2023} challenges. As the DeepVQE model processes audio sampled at 24 kHz, we downsample the blind test set audio and upsample the result as described in Section \ref{feat_extraction}.

\noindent \textbf{Experimental setup.}
To test the echoes removal, we employ the echo return loss enhancement (ERLE) for far end single-talk (FEST) scenarios and AECMOS \cite{purin2022aecmos} echo score for both FEST and DT. For near end single-talk (NE) scenarios, we use the word error rate (WER), AECMOS \cite{purin2022aecmos} degradation score, DNSMOS P.835 signal (SIG), background (BAK), overall (OVRL) scores \cite{reddy2022dnsmos} and the speech-to-reverberation-ratio (SRR), estimated with an internal model. In the NS scenarios, we feed into the network an empty far end signal.

\noindent \textbf{Hyper-parameters.}
For feature generation, we used a squared root Hann window of length $20$ms, a hop length of $10$ms, and a discrete Fourier transform length of 480. This leads to $20$ms algorithmic delay, consisting of $10$ms output signal delay (due to overlap-add signal reconstruction) and packet (frame) duration ($10$ms). We trained all the networks using AdamW \cite{loshchilovdecoupled} optimizer with batches of $400$ samples for $250$ epochs, with a learning rate of $1.2 \cdot 10^{-3}$ and a weight decay of $5 \cdot 10^{-7}$. Similar to \cite{indenbom2022deep}, we have taken $d_{max}=100$, which is equivalent to the maximum delay of $1$ second.

\noindent \textbf{Ablation study.}
We performed the ablation studies with the small configuration of our DeepVQE model, named DeepVQE-S. DeepVQE-S is a downscaled version of our best model. The DeepVQE-S microphone branch has 4 blocks with 16, 40, 56, and 24 filters, the far end branch has 8 and 24 filters, and the decoding branch has 4 blocks with 40, 32, 32, and 27 filters. Additionally, the residual block is omitted in all the encoder blocks and in the first and last decoder blocks to save more computing. We find it is more interesting to show ablation results on the production-sized model, where the quality impact of architectural changes is especially important.

In Table \ref{tab_results_align}, we compare our alignment block against the DSP-based alignment method (the DeepVQE-S without the alignment block) and the alignment block proposed in \cite{indenbom2022deep}. We observe that our method surpasses both methods in each and every metric with a considerably higher improvement obtained for WER. We highlight that all the compared models are derived from the DeepVQE-S model and include residual blocks, sub-pixel convolutions, CCM blocks, etc. An extended ablation for the AEC task with comparative audio samples and visualizations of the alignment delay map is presented in \url{https://ristea.github.io/deep-vqe}.

In Table \ref{tab_results_ns}, we present the ablation results for the NS task on both 2022 DNS Challenge \cite{dubey2022icassp} and 2023 AEC Challenge \cite{AEC-2023} blind test sets. We remove or replace each of the proposed blocks to see the impact of each architectural change separately. The biggest improvement is provided by the CCM block, showing the potential of utilizing magnitude and phase information from neighboring frequency bins and preceding time frames. In terms of DR, our best DeepVQE-S shows over $10$dB SRR improvement on both AEC-NEST and DNS data, showing a great capacity to jointly perform AEC, NS and DR.

\noindent \textbf{Challenge results.}
In Table \ref{tab_results_DNSmos} and Table \ref{tab_results_AECmos} we included the subjective mean opinion score (MOS), word accuracy rate (WAcc), and the overall final score from the 2023 AEC \cite{AEC-2023} and DNS \cite{DNS-2023} challenges for our best DeepVQE model. Considering that our model does not use personalized information, we compared ourselves with the non-personalized models. We significantly overpass the winner of the DNS Challenge for both tracks, obtaining better results for 3 out of 4 metrics. Regarding the AEC Challenge winner, we obtained better metrics for ST NE scenarios, while being slightly behind for ST FE and DT scenarios. Nevertheless, according to the official final score, we rank first.

We highlight that the winners from AEC and DNS challenges are different models specifically designed and trained for the challenge task, while DeepVQE is exactly the same model, trained to jointly perform both tasks.

\noindent \textbf{Inference speed.}
Providing SOTA performance within a speed-parameters budget is critical to deploy the models in production for teleconferencing applications. DeepVQE has $7.5$M parameters and an inference time of $3.66$ms per frame on a CPU Intel Core i7 11370H@3.3 GHz, while DeepVQE-S has only $0.59M$ parameters and an inference time of $0.14$ms per frame. Having a very good performance and a real-time factor of $0.014$, DeepVQE-S has been successfully tested in Microsoft Teams for hundreds of millions of users.

\vspace{-0.2cm}
\section{Conclusions}
\vspace{-0.1cm}
In this paper, we propose the DeepVQE architecture, a new model for real-time unified AEC, NS, and DR. Our model contains a more stable alignment block, which significantly improves the AEC performance. Moreover, the model contains residual blocks, sub-pixel convolutions, and CCM blocks, which enables the model to attain SOTA performance in NS. DeepVQE model achieved SOTA performance in both 2023 AEC \cite{AEC-2023} and 2023 DNS \cite{DNS-2023} challenges, which leads the way to unified speech modeling. In the future, we aim to improve the performance of the proposed model and extend the architecture for personalized AEC and NS.





\bibliographystyle{IEEEtran}
\bibliography{mybib,IC3-AI}

\end{document}